\documentstyle[preprint,aps,epsfig]{revtex}

  \usepackage{epsfig}
  \usepackage{amsmath}
  \usepackage{times}

\def\be{\begin{equation}}    
\def\ee{\end{equation}}
\def\lk{\left(}
\def\rk{\right)}
\def\dmin{d_{{\rm min}}}
\def\dmax{d_{{\rm max}}}

\begin{document}
\draft

\title{Magnetization of polydisperse colloidal ferrofluids: \\ Effect of magnetostriction}


\author{J. P. Huang and C. Holm}
\address{Max Planck Institute for Polymer Research, Ackermannweg 10, 55128, Mainz, Germany}


\date{\today}

\maketitle

\begin{abstract}
We  exploit magnetostriction in polydisperse ferrofluids in order to 
generate  nonlinear responses, and apply a thermodynamical method to 
derive the desired nonlinear magnetic susceptibility. For an ideal gas, 
this method has been demonstrated to be in excellent agreement with a 
 statistical method. In the presence of a sinusoidal ac 
magnetic field, the magnetization of the polydisperse ferrofluid   
contains higher-order harmonics, which can be extracted analytically 
by using a perturbation approach. We find that the harmonics are 
sensitive to the particle distribution  and the degree of field-induced anisotropy 
of the system. In addition, we  find that the magnetization is higher 
in the polydisperse system than in the  monodisperse one, as also 
found by a recent Monte Carlo simulation. Thus, it seems 
possible to detect the size distribution   
in a polydisperse ferrofluid by measuring the harmonics of the 
magnetization under the influence of magnetostriction.
\end{abstract}





\newpage

\section{Introduction}

Ferrofluids (magnetic fluids) are colloidal suspensions containing
single domain nanosize ferromagnetic particles dispersed in a
carrier liquid~\cite{r1}. These particles are usually stabilized
against agglomeration by coating them with long-chain
molecules (steric stabilization), or decorating them with charged groups
(electrostatic stabilization). Since these particles can 
easily interact via
applied magnetic fields, which in turn can affect the viscosity
and structural properties, ferrofluids possess a wide
variety of potential applications in many fields ranging from
mechanical engineering~\cite{Ber,Od} to biomedical 
applications~\cite{Hergt,Al}. Thus, ferrofluids have  received much 
attention in the scientific 
community~\cite{Liu1,Liu2,r2,r3,r4,r5,r6,Luke,r7,r8,Patrick,r9,r10,Kri,AR1}.

Polydispersity of ferrofluids emerges naturally since the
particles in them always possess a log-normal
distribution~\cite{r2,r3,r4}. 
It has been experimentally observed that
different microstructures can spontaneously form in ferrofluids~\cite{r4}. 
This has a strong effect on the macroscopic properties of ferrofluids. In
this regard, the influence of polydispersity on the
magnetization of ferrofluids is of both
academic and practical interest. Due to the 
polydispersity of ferrofluids, their structure and magnetization properties 
may significantly differ from those of monodisperse
systems~\cite{r5,r6,Luke,r7}.  

The structure of polydisperse
ferrofluids has been discussed theoretically on the basis of a
bidisperse model in which the fluids consist of two fractions of
magnetic particles with significant size differences~\cite{r8,r9}.
In a bidisperse ferrofluid, the smaller particles are affected
by Brownian motion, and are therefore more or less randomly dispersed. The larger
magnetic dipole moment of the larger particles leads, however, 
to  a strong interparticle force which dominates over Brownian motion. Thus the
salient structure in these systems is proposed to be a chainlike
aggregate formed by the larger particles. Some small particles
may be attached to the ends of these aggregates, but most of
them remain in the non-aggregated state~\cite{r9}. These 
features have also been observed in Langevin dynamics
simulations~\cite{r10}.
In addition, Wang and Holm~\cite{r10} found that the smaller
particles hinder the aggregation of larger particles. Since then, 
that effect has been  investigated in detail by
Zubarev~\cite{Zubarev04}, and  the influence 
of polydispersity on the equilibrium properties of ferrofluids was 
very recently investigated by Krist\'{o}f and Szalai using 
Monte Carlo simulations~\cite{Kri}. 
Krist\'{o}f and Szalai found that magnetization is generally higher in a 
polydisperse system than in the corresponding monodisperse one.

An inhomogeneous magnetic field ${\bf H}$ exerts 
a translational force ${\bf F}$ on a ferromagnetic particle  given by
\be
{\bf F} = \alpha {\bf H}\cdot \nabla {\bf H}+{\bf m}_0\cdot \nabla {\bf H},
\ee
where $m_0$ and $\alpha$ are  the permanent magnetic dipole moment  
and the magnetizability of the particle, respectively. Thus, if the 
permanent moment points in the direction of ${\bf H}$, the particles 
will be displaced towards the regions of higher field strength. In a 
macroscopic sample the average moment is in the direction of the field, i.\,e.,
the particles favor orientations where their permanent magnetic 
dipole moments are directed along the field. Thus, an inhomogeneous 
field acting on a macroscopic sample causes a concentration gradient 
with high concentrations at high field strengths. If a sample 
is situated partially in a field and held at constant pressure, 
the particle density in the field region will increase leading to 
an increase of the permeability. This effect is called magnetostriction, 
or in general, a response of the solution to an inhomogeneous magnetic field.  
Magnetostriction has been extensively studied, 
e.\,g., in dipolar fluids~\cite{Bot}, single crystals of the high-$T_c$ 
cuprate ${\rm Bi}_2{\rm Sr}_2{\rm CaCu}_2{\rm O}_8$~\cite{Ikuta}, 
polycrystalline Fe films~\cite{Weber}, and in cylindrical type-II 
superconductors~\cite{Joh}. Unfortunately, to the best of 
our knowledge, so far there is  neither theory nor experiments
dealing with the important problem of magnetostriction of ferrofluids.
The only notable exception is ferrogels~\cite{EJ2003} which are 
chemically cross-linked  polymer networks swollen with a ferrofluid. 

To model ferrofluids we use a thermodynamical method to 
derive the magnetostriction-induced effective third-order  nonlinear  
magnetic susceptibility  $\xi$. As a sinusoidal 
ac magnetic field is applied, the magnetization will, in general, consist 
of ac fields at frequencies of the higher-order harmonics. 
We derive the harmonics analytically by using 
a perturbation approach. The aim of the present paper is to exploit 
magnetostriction in ferrofluids in order to generate a nonlinear
response and thus a harmonic response.  
In experiments, measurements of the harmonic responses of magnetization
have been used to obtain information of the anisotropy 
distribution in a ferromagnetic amorphous alloy~\cite{AG92}.


This paper is organized as follows. In Sec.~II, the nonlinear magnetic 
susceptibility arising from the magnetostriction is derived, 
and the harmonics of the magnetization 
are extracted analytically. In Sec.~III, we numerically 
calculate the harmonics of the magnetization under various conditions. 
We finish with a discussion and conclusions in Sec.~IV.

\section{Formalism}

\subsection{Nonlinear characteristic arising from magnetostriction}

Let us assume that a ferrofluid is placed in an inhomogeneous magnetic 
field and kept at constant pressure. Then, the density of the particles in 
the field regions will increase due to the interaction between
the permanent magnetic moments of the particles and the field 
leading to an increase in the permeability.
This effect is called magnetostriction. 

The experimental situation is the following: There is a field-affected 
volume  with volume $V_c ,$ in which the magnetic field and the magnetic 
induction are denoted by $H_c$ and $B_c,$ respectively. Both of them 
should satisfy the usual magnetostatic equations~\cite{B62},
\begin{eqnarray}
\nabla \cdot {\bf B}_c &=& 0,\\
\nabla\times {\bf H}_c &=& 0.\label{Ec}
\end{eqnarray}
Here Eq.~(\ref{Ec}) implies that the magnetic field ${\bf H}_c$ can 
be expressed as the gradient of a magnetic scalar potential $\Phi$ such that
\be
 {\bf H}_c = -\nabla\Phi.
\ee
Under appropriate boundary conditions, the inhomogeneous ferrofluid inside 
the  field-affected volume can be represented as a region of volume $V_c,$ 
surrounded by a surface $S'$. Such boundary conditions can be written as
\be
\Phi = - {\bf H}\cdot {\bf X}\,\,{\rm on}\,\,S',
\ee
which, if the ferrofluid within $V_c$ were uniform, would give rise to 
a magnetic field which is identical to ${\bf H}$ (external field) 
everywhere within $V_c.$ In fact, this boundary condition guarantees 
that even in an inhomogeneous ferrofluid the volume average of the magnetic 
field $\langle {\bf H}_c\rangle$ within $V_c$ still equals that of 
the external field, $\langle {\bf H}\rangle$, namely,
\be
\langle {\bf H}_c\rangle = \frac{1}{V_c}\int {\bf H}_c({\bf X}){\rm d}^3x = \langle {\bf H}\rangle.
\ee
In this case  there is no external field outside the  field-affected volume and
(or, in practice, the external field in other areas is  weak enough to be neglected). 
The ferrofluid with volume $V$ is situated such that it has regions 
both inside and outside the field-affected volume at a constant pressure~$p$.

In the presence of an inhomogeneous external magnetic field ${\bf H}$, 
the effective linear permeability $\mu_e$ and the effective third-order 
nonlinear magnetic susceptibility $\xi$ for the ferrofluid inside 
the  field-affected volume are defined as
\be
\langle {\bf B}_c\rangle 
= \left (\mu_e +4\pi\xi \langle{\bf H}\rangle ^2\right )\langle{\bf H}\rangle ,
\label{delf1}
\ee
where
$\langle\cdots\rangle$ denotes the volume average. 
Eq.~(\ref{delf1}) implies that there is a nonlinear relation 
between the magnetization, $M$, and the magnetic field, $\langle{\bf H}\rangle$. 
This will be explicitly shown later in Eq.~(\ref{Mor}). 
Further, it is worth noting that 
the nonlinear term $\xi$ of the magnetic susceptibility is actually 
an effective quantity. It appears due to the fact that the
particles can receive a translational force which drives them into 
the field-affected volume in the presence of an inhomogeneous field. 
In this regard, rather than ${\bf H}$ we should use $\langle{\bf H}\rangle$ 
in Eq.~(\ref{delf1}) 
in order to derive the (effective) nonlinear term $\xi$.



Alternatively, based on thermodynamics, the permeability $\mu_H$ including the 
incremental part  due to the magnetostriction can be defined as
\be
\mu_H =\lk\frac{\partial \langle {\bf B}_c\rangle }
{\partial \langle {\bf H}\rangle }\rk_{T,p}
= \lk\frac{\partial \langle {\bf B}_c\rangle }
{\partial \langle {\bf H}\rangle }\rk _{T,\rho}
+\int_{d_{{\rm min}}}^{d_{\rm max}}f(d)\lk
\frac{\partial \langle {\bf B}_c\rangle }
{\partial\rho (d)} \rk_{T,\langle {\bf H}\rangle }\lk
\frac{\partial\rho (d)}{\partial 
\langle {\bf H}\rangle } \rk_{T,p}{\rm d}d,
\label{muH}
\ee 
where $\rho$ stands for the density of the part of the 
ferrofluid inside the  field-affected volume, and
$\dmin$ (or $\dmax$) denotes the minimum (or maximum) 
particle diameter. Here the size distribution of particles 
$f(d)$ satisfies the known log-normal law~\cite{r3,Payet} 
\be
f(d)=\frac{1}{\sqrt{2\pi}\sigma d}
\exp [-\frac{\ln^2 (d/\delta)}{2\sigma^2}],
\label{lnD}
\ee
where $\sigma$ is the standard deviation of $\ln d$ and $\delta$ the median diameter.  

In Eq.~(\ref{muH}),  
$\lk\frac{\partial \langle {\bf B}_c\rangle }
{\partial \langle {\bf H}\rangle }\rk _{T,\rho}$ 
represents the effective permeability including all nonlinear effects at a constant density.
As a matter of fact, regarding both Eqs.~(\ref{delf1})~and~(\ref{muH}), 
the total effective third-order nonlinear susceptibility generally 
contains two contributions. The first is the magnetostriction-induced one
considered in this work, and the other is the normal-saturation contribution. 
At large field intensities, the higher terms of the Langevin function should 
be taken into account. This contribution is negative, and is called 
normal saturation. In contrast, the magnetostriction has a positive contribution.
In this work, we assume that the field is moderate such 
that the normal-saturation contribution is weak enough to be neglected. 
It should be noted that the argument of the Langevin function can become large for a 
very small number of large particles in the tail of the size distribution. 
Further, we shall also neglect the normal-saturation contribution resulting 
from the very small amount of the large particles since this contribution 
might be expected to have a very weak effect on the total effective third 
order susceptibility. To summarize the above, throughout this work, 
only the magnetostriction-induced  contribution is considered. 

Using the above assumptions
$\lk\frac{\partial \langle {\bf B}_c\rangle }
{\partial \langle {\bf H}\rangle }\rk _{T,\rho} \equiv \mu_e$ 
can be given by the anisotropic  Clausius-Mossotti equation, namely~\cite{LoPRE01}
\be
\frac{g_L(\mu_e-\mu_2)}{\mu_2+g_L(\mu_e-\mu_2)}
=\frac{4\pi}{3}\int_{\dmin}^{\dmax}f(d)N(d)
\left (\alpha (d)+\frac{m_0(d)^2}{3k_BT}\right ){\rm d}d,
\label{CME}
\ee
where $\mu_2$ denotes the permeability of the carrier liquid, $m_0(d)$ [or $\alpha(d)$] 
the permanent magnetic dipole moment (or magnetizability) of a particle
with diameter $d$, $N(d)$ is the number density of particles with diameter 
$d$, $k_B$ the Boltzmann constant, and $T$ the temperature. 
Regarding Eq.~(\ref{CME}),  more issues should be remarked. 
It is known that the 
usual (isotropic) Clausius-Mossotti equation does not include the 
particle-particle interaction. When Lo and Yu studied the field-induced 
structure transformation in electrorheological solids, they succeeded in 
developing a generalized Clausius-Mossotti equation by introducing 
a local-field factor $\beta '$ which reflects the particle-particle 
interaction between the particles in an anisotropic lattice. This
generalized Clausius-Mossotti approach  [Eq.~(\ref{CME})] is a 
self-consistent determination of the local  field due to a lattice 
of dipole moments. In other words, Eq.~(\ref{CME})  should be expected to 
include particle-particle interactions (at least to some extent),
and the degree of the particle-particle interactions depends on how 
much $g_L$ deviates from $1/3$ (note that here  $g_L=\beta'/3$),
 where $g_L$ represents the demagnetizing 
factor in the longitudinal field case. In particular, the case when 
$1/g_L=3$  (or $g_L=1/3$) 
corresponds to the isotropic case, which yields the well-known (isotropic) 
Clausius-Mossotti equation. It is worth noting that there is a sum rule 
$g_L+2g_T=1$~\cite{Lan}, where $g_T$ is the demagnetizing factor in the 
transverse field case. For electrorheological fluids, similar factors 
were measured in simulations~\cite{Martin98,Martin2}.  

In Eq.~(\ref{CME}), 
the term $\frac{m_0(d)^2}{3k_BT}$ results from the average contribution 
of the permanent magnetic dipole moment to the average value of the 
work required to bring a particle with diameter $d$ into the field 
$\langle {\bf H} \rangle$. More precisely, the mean value of the 
component of the dipole moment in the direction of the field is given by
\be
m_0(d)L(\gamma) = \frac{m_0(d)^2}{3k_BT}\langle  H\rangle,
\ee
with $\gamma = \frac{m_0(d)\langle H\rangle }{k_BT}$. That is, we 
set the Langevin function to $L(\gamma)=\gamma/3$. Regarding this 
relation, the following issues should be noted. In the present work, we shall adopt
a perturbation approach~\cite{Gu92} [see Sec.~II(C)], which is suitable 
for a weak nonlinearity.
In the perturbation approach, it is well established that the effective
third-order nonlinear susceptibility  can be calculated from the linear 
field~\cite{Stroud88}, while the effective higher-order  nonlinearity must depend on
the nonlinear  field~\cite{Gu92}. In fact, we could have adopted a
self-consistent mean-field approach~\cite{PRE1}, but the perturbation approach
appears to be more convenient for analytic expressions~\cite{PRE1}. 
Thus, to be able to focus on (weak) third-order nonlinearity, it suffices  to use the Clausius-Mossotti
equation [Eq.~(\ref{CME})] by taking into account  the linear relation only, 
i.\,e.,  $L(\gamma)=\gamma/3.$ Due to the same reason, in what follows 
we shall omit the contribution from the  nonlinear  field, too.

Regarding the incremental permeability due to the magnetostriction, namely the last term in Eq.~(\ref{muH}) 
$$
\int_{d_{{\rm min}}}^{d_{\rm max}}f(d)\lk
\frac{\partial \langle {\bf B}_c\rangle }
{\partial\rho (d)} \rk_{T,\langle {\bf H}\rangle }\lk
\frac{\partial\rho (d)}{\partial \langle {\bf H}\rangle } 
\rk_{T,p}{\rm d}d\equiv 12\pi\xi\langle{\bf H}\rangle ^2,
$$
we obtain
\be
12\pi\xi\langle{\bf H}\rangle ^2 
=  \int_{\dmin}^{\dmax}f(d)\langle {\bf H}\rangle \lk\frac{\partial\mu_e}
{\partial\rho (d)}\rk_{T,\langle {\bf H}\rangle }\lk\frac{\partial\rho (d)}
{\partial \langle {\bf H}\rangle }\rk_{T,p}{\rm d}d.
\label{mue}
\ee
This equation is valid for the lowest-order perturbation. The differential increase 
of the density inside the  field-affected volume  ${\rm d}\rho(d)$ corresponds 
to an increase in mass given by $V_c{\rm d}\rho(d)$. This increase is
equal to the decrease in mass outside the  field-affected volume  given by 
$-\rho(d) {\rm d}(V-V_c)=-\rho (d) {\rm d}V,$ so that ${\rm d}\rho(d)=-[\rho(d)/V_c]{\rm d} V$. 
Thus, we may rewrite Eq.~(\ref{mue}) as
\be
12\pi\xi\langle{\bf H}\rangle ^2 
= -\int_{\dmin}^{\dmax}f(d)\langle {\bf H}\rangle \lk
\frac{\partial\mu_e}{\partial\rho(d)}\rk_{T,\langle {\bf H}\rangle}
\frac{\rho(d)}{V_c}\lk
\frac{\partial V}{\partial \langle {\bf H}\rangle }\rk_{T,p}{\rm d}d.
\label{mue2}
\ee

Next, we can obtain 
$\lk\frac{\partial V}{\partial \langle {\bf H}\rangle }\rk_{T,p}$ 
based on the differential of the free energy ${\rm d}F$:
\be
{\rm d}F=-S{\rm d} T-p{\rm d}V+\frac{V_c}{4\pi}\langle {\bf H}\rangle {\rm d}\langle {\bf B}_c\rangle ,
\ee
where $S$ denotes the entropy.
Introducing the transformed free enthalpy $\bar{G}$,
\be
\bar{G} = F+pV-\frac{V_c}{4\pi}\langle {\bf H}\rangle \langle {\bf B}_c\rangle ,
\ee
we obtain the following expression for its differential:
\be
{\rm d}\bar{G} = -S{\rm d}T+V{\rm d}p-\frac{V_c}{4\pi}\langle {\bf B}_c\rangle {\rm d}\langle {\bf H}\rangle .
\ee
From this equation, we find
\be
\lk\frac{\partial V}{\partial \langle {\bf H}\rangle } \rk_{T,p} 
= -\frac{V_c}{4\pi}\lk
  \frac{\partial \langle {\bf B}_c\rangle }{\partial p}\rk_{T,\langle {\bf H}\rangle } 
= -\frac{V_c\langle {\bf H}\rangle }{4\pi}\lk
\frac{\partial\mu_e}{\partial p} \rk_{T,\langle {\bf H}\rangle }.
\ee
Then, the substitution of this into Eq.~(\ref{mue2}) leads to
\be
12\pi\xi \langle{\bf H}\rangle ^2 
= \int_{\dmin}^{\dmax}f(d)\frac{\langle {\bf H}\rangle ^2}{4\pi}\rho(d)\lk
\frac{\partial\mu_e}{\partial \rho(d)}\rk_{T,\langle {\bf H}\rangle }\lk
\frac{\partial\mu_e}{\partial p}\rk_{T,\langle {\bf H}\rangle }{\rm d}d.
\label{muee}
\ee

We now use
\be
\lk\frac{\partial\mu_e}{\partial p}\rk_{T,\langle {\bf H}\rangle }
=\lk\frac{\partial\mu_e}{\partial\rho(d)}\rk_{T,\langle {\bf H}\rangle }\lk
\frac{\partial\rho(d)}{\partial p} \rk_{T,\langle {\bf H}\rangle }
=\beta\rho(d)\lk\frac{\partial\mu_e}{\partial\rho(d)}\rk_T,
\label{compre}
\ee
where $\beta=-\frac{1}{V}\lk\frac{\partial V}{\partial p}\rk_T$ is 
the compressibility in the absence of the field. In the last term 
of Eq.~(\ref{compre}), terms depending on $\langle {\bf H}\rangle $ 
have been neglected since they lead to terms in powers of 
$\langle {\bf H}\rangle $  higher than the second term in Eq.~(\ref{muee}). 
In the same approximation, after substitution of Eq.~(\ref{compre}) into  Eq.~(\ref{muee}) 
we obtain
\be
12\pi\xi\langle{\bf H}\rangle ^2  
= \int_{\dmin}^{\dmax}f(d)\frac{\langle {\bf H}\rangle ^2}{4\pi}\beta\rho(d)^2\lk
\frac{\partial\mu_e}{\partial \rho(d)}\rk^2 _T{\rm d}d.
\ee
Thus, we have
\be
\xi = \frac{\beta}{48\pi^2}\int_{\dmin}^{\dmax}f(d)\rho(d)^2\lk
\frac{\partial\mu_e}{\partial \rho(d)}\rk^2 _T{\rm d}d.
\label{xi-final}
\ee
So far, the effective third-order nonlinear magnetic susceptibility $\xi$ has
been derived in terms of the compressibility, size distribution function, and the 
differential of effective linear permeability with respect to the density.

\subsection{Comparison with a  statistical method}

The increase of the density $\Delta\rho$ due to magnetostriction can be 
calculated from $\lk \partial V/\partial \langle {\bf H}\rangle \rk_{T,p}$.  
Details are as follows:
\begin{eqnarray}
\Delta\rho &=& \int_{\dmin}^{\dmax}\int_0^{\langle {\bf H}\rangle} f(d)\lk
\frac{\partial\rho(d)}{\partial \langle {\bf H}\rangle }\rk_{T,p} 
{\rm d}\langle {\bf H}\rangle {\rm d}d \nonumber \\
&=& \int_{\dmin}^{\dmax}\int_0^{\langle {\bf H}\rangle} 
-f(d)\frac{\rho(d)}{V_c}\lk\frac{\partial V}
{\partial \langle {\bf H}\rangle } \rk_{T,p} {\rm d}\langle {\bf H}\rangle {\rm d}d \nonumber \\
&=& \int_{\dmin}^{\dmax}\int_0^{\langle {\bf H}\rangle} 
\frac{\langle {\bf H}\rangle \rho(d)}{4\pi}\lk\frac{\partial\mu_e}{\partial p} 
\rk_{T,\langle {\bf H}\rangle } {\rm d}\langle {\bf H}\rangle {\rm d}d \nonumber \\
&=& \int_{\dmin}^{\dmax}\int_0^{\langle {\bf H}\rangle} 
\frac{\langle {\bf H}\rangle \beta\rho(d)^2}{4\pi}\lk\frac{\partial\mu_e}
{\partial\rho(d)} \rk_T {\rm d}\langle {\bf H}\rangle {\rm d}d.\label{density}
\end{eqnarray}
To obtain this equation, Eq.~(\ref{compre}) has been used, which means that 
in the expression for $\Delta\rho$ terms in powers of $\langle {\bf H}\rangle $ 
higher than the second have been neglected. Then, the integration can be performed, and we obtain
\be
\Delta\rho = \int_{\dmin}^{\dmax}\frac{\langle {\bf H}\rangle ^2}{8\pi}\beta \rho(d)^2\lk\frac{\partial\mu_e}{\partial \rho(d)}\rk_T {\rm d}d.
\label{delRho}
\ee

For comparing with a statistical method, let us apply our formalism 
to a (monodisperse) ideal gas~\cite{Bot}. For its effective permeability, 
in view of $g_L=1/3$,   Eq.~(\ref{CME}) can be rewritten as 
\be
\frac{\mu_e-1}{\mu_e+2}=\frac{4\pi}{3}N\left (\alpha+\frac{m_0^2}{3k_BT}\right ).\label{CME2}
\ee
For the ideal gas, $\mu_e\sim 1,$ and hence we obtain
\be
\lk\frac{\partial\mu_e}{\partial\rho} \rk_T=\frac{4\pi}{m_m}\lk \alpha+\frac{m_0^2}{3k_BT} \rk ,
\ee
where $m_m$ denotes the mass of a single molecule, 
$m_m=M/N_A$. Here $M$ is the molecular weight 
and $N_A$ the Avogadro constant. For this case, using $\beta=M/\rho R T,$ Eq.~(\ref{delRho}) predicts
\be
\Delta\rho = \frac{N_A\langle {\bf H}\rangle^2 \rho}{2RT}\lk \alpha+\frac{m_0^2}{3k_BT}\rk .\label{Gas}
\ee
Here $R$ represents the molar gas constant.

This result [Eq.~(\ref{Gas})] can also be achieved by using a statistical 
method. Because of Boltzmann's distribution law, $n^{\langle {\bf H}\rangle}$, 
the number of moles per cm$^3$ of the gas at a point with field strength 
$\langle {\bf H}\rangle$, is given by
\be
n^{\langle {\bf H}\rangle} = n^{(0)}\exp(-W/k_BT),
\ee
where $n^{(0)}$  is the number of moles per cm$^3$ of the gas at a point with zero 
field, and $W$ the average value of the work required to bring a molecule into 
the field  $\langle {\bf H}\rangle$, 
\be
W = -{1 \over 2} \lk \alpha+\frac{m_0^2}{3k_BT}\rk \langle {\bf H}\rangle^2 .
\ee
Thus, we have 
\be
n^{\langle {\bf H}\rangle} = n^{(0)}\exp\left [ {1 \over 2} \lk 
\alpha+\frac{m_0^2}{3k_BT}\rk \langle {\bf H}\rangle^2 /k_BT \right ].
\ee
Let us neglect the terms in higher than second powers of  $\langle {\bf H}\rangle$, and we have
\be
\Delta\rho = M(n^{\langle {\bf H}\rangle} - n^{(0)})
=\frac{N_A\langle {\bf H}\rangle^2 \rho}{2RT}\lk \alpha+\frac{m_0^2}{3k_BT}\rk .\label{Bolt}
\ee
For an ideal gas, Eq.~(\ref{Bolt}) yields exactly the same result as Eq.~(\ref{Gas}), 
albeit derived using a different approach. This shows the consistency of our arguments.

\subsection{Magnetization and high-order harmonics}

The orientational magnetization ($z$ axis) has the following general form
\be
{\bf M} = \frac{\mu_e-\mu_2}{4\pi}\langle {\bf H}\rangle +\xi \langle {\bf H}\rangle ^3.
\label{Mor}
\ee
Here, the higher order terms have been omitted.
We use an inhomogeneous sinusoidal ac field
${\bf H} = (l_z/L_z) H_{\rm ac}(t)\hat{{\bf z}} 
=(l_z/L_z) H_{{\rm ac}}\hat{{\bf z}}\sin(\omega t)$, 
where $0< l_z\le L_z$ with $L_z$ being the length of the  field-affected volume   
along $z$ axis. Without the loss of generality, we set $L_z=1$ in the following.
We now obtain
\be
M = \frac{\mu_e-\mu_2}{8\pi}H_{{\rm ac}}(t)+\frac{\xi}{8} H_{{\rm ac}}(t)^3.
\ee
In view of $H_{{\rm ac}}(t) = H_{{\rm ac}}\sin(\omega t)$, 
the magnetization $M$ can be expressed in terms of the odd-order harmonics such that
\be
M = M_{\omega}\sin(\omega t) + M_{3\omega}\sin (3\omega t),
\ee
where the fundamental and third-order harmonics are given by
\begin{eqnarray}
M_{\omega} &=&  \frac{\mu_e-\mu_2}{8\pi}H_{{\rm ac}}+\frac{3\xi}{32}H_{{\rm ac}}^3,\\
M_{3\omega} &=& -\frac{\xi}{32}H_{{\rm ac}}^3.
\end{eqnarray}
In the above derivation, we have used the identity 
$\sin^3 (\omega t) = (3/4)\sin(\omega t) - (1/4) \sin (3\omega t)$.

\section{Numerical results}

Without any loss of generality, we choose the following parameters
for our numerical calculations: 
$\mu_2=1$ (non-magnetic carrier fluid),   density of the bulk 
material of the particles $7$\,g/cm$^3$,  $H_{{\rm ac}}=20$\,Oe, 
$\beta = 0.62\times 10^{-10}$\,cm$\cdot$s$^2$/g, $\alpha(d)=0$ 
(due to  the small size of the particles), and  $T=298$\,K. In addition, 
the volume fraction of the particles   
is set to be $0.08$, and the saturation magnetization of the bulk 
material of the particles is $450$\,emu.  
Finally, setting $\dmin=1\,$nm, and $\dmax=30\,$nm ensures
$$
\int_{\dmin}^{\dmax}f(d){\rm d}d \simeq 1
$$
as expected. 

Based on the model parameters, we calculated the dipolar coupling 
constant~\cite{W-PRE02} 
$\lambda (\delta) = \frac{m_0(\delta)^2}{\mu_0k_BT\delta^3}$, 
and found 
$\lambda(9.5\,{\rm nm})=1.16,  \lambda(10\,{\rm nm})=1.35,  \lambda(10.5\,{\rm nm})=1.56$, 
which ensures the assumption that the particle interaction in our system is weak.

In Fig.~1, we display the size distribution of the particles in the 
lognormal law for different (a) median diameter $\delta$ and (b) standard deviation $\sigma$.  
Fig.~2 shows the fundamental [Fig.~2(a)] and third-order [Fig.~2(c)] harmonics 
of the magnetization as a function of the degree of anisotropy $1/g_L$ for 
different median diameters $\delta$. The size distribution 
of the particles is shown in Fig.~1(a). It is found that increasing the 
degree of anisotropy $1/g_L$ causes both the fundamental and third-order 
harmonics to increase. Also, a higher median diameter $\delta$ leads to 
larger harmonics.

In Fig.~3, the fundamental [Fig.~2(b)] and third-order [Fig.~2(c)] 
harmonics of the magnetization are plotted as a function of $1/g_L$ 
for different standard deviations $\sigma$. The 
lognormal size distribution of the particles is  shown in Fig.~1(b). 
Again, it is shown that  the  harmonics increase 
with increasing median diameter $\delta$.

Finally, to compare the above polydisperse case with the corresponding 
monodisperse one, we study the monodisperse case in Fig.~4 for three 
different diameters which have the same values as the median diameters 
used in Fig.~2. In the monodisperse case, it is also 
evident that increasing the degree of anisotropy $1/g_L$ causes both 
the fundamental and third-order harmonics to increase. In addition, larger  
diameter leads to larger harmonics. It is worthing noting that both 
the fundamental and third-order harmonics of the magnetization are higher 
in the polydisperse system than in the monodisperse one when comparing 
Fig.~2 with Fig.~4. In particular, the third-order harmonics of the 
polydisperse system can be of two orders of magnitude larger than 
those of the monodisperse system. In other words, the magnetization 
is higher in the polydisperse system than in the monodisperse one, 
due to the fact that for this comparison the volume fraction of the 
particles is fixed. This is in agreement with 
the findings of Ref.~\cite{Kri} where a Monte Carlo 
simulation was used to study the influence of polydispersity on 
the equilibrium properties of ferrofluids.

\section{Discussion and conclusion}


Here some comments are in order. In the present paper, we have exploited 
magnetostriction in ferrofluids in order to generate  nonlinear responses. 
The proposed mechanism should  work for dc magnetic fields. It will also work for ac fields 
with  frequency $\nu = \omega/(2\pi)$ if the size of the sample 
is not greater than $c_{s}/\nu$, where $c_{s}$ is the sound velocity. Thus, 
$\nu$ can be up to kHz or so. Otherwise the required mass density oscillations 
will not be able to keep up with the rapid changes in the magnetic field. 

To obtain the lowest-order (i.\,e., cubic) nonlinearity, we have 
assumed that material properties such as permeability of the polydisperse 
system can be calculated as a linear superposition of the corresponding values 
in the monodisperse systems, see Eqs.~(\ref{muH}),~(\ref{CME})~and~(\ref{density}). 
For Eq.~(\ref{muH}), the linear superposition  should hold since the nonlinear 
term $\xi$ is actually an effective quantity which results from all the 
monodisperse systems. For Eq.~(\ref{CME}), we used the linear superposition 
again. The reason is that the right-hand side of Eq.~(\ref{CME}) actually 
represents the effective contribution from two parts: The induced magnetization 
(which has been assumed to disappear due to the small size of the particles 
in our numerical calculations), and the permanent-moment-related magnetization. 
In addition, once the inhomogeneous field is applied, the particles with 
different sizes are able to move into the field-affected volume, thus yielding an 
increasing particle density. In this regard, for Eq.~(\ref{density}), 
the linear superposition should be used naturally.

Nonlinear optical materials  with large values of (effective) third-order
nonlinear dielectric susceptibilities~\cite{Smith84} are in great need
in industrial applications such as nonlinear optical switching
devices for use in photonics, and real-time coherent optical signal
processors, and so on. Due to the similarity between magnetics and dielectrics,  
the present (effective) third-order nonlinear magnetic susceptibilities 
are expected to have some potential applications in nonlinear magnetic  devices. 



Throughout the paper, only odd-order harmonics are induced to appear. 
As a matter of fact, if one applies an ac magnetic field superimposed 
to a dc field, the even-order harmonics should appear~\cite{AG92}.
That is due to the coupling between the two kinds of fields. On 
the other hand, since the second-order harmonics are often of 
several orders of magnitude larger than the corresponding third-order, 
the second-order harmonics are more attractive for the experimental measurements~\cite{AG92}.

We have considered the fundamental and third-order harmonics. In fact, 
we can consider much higher-order harmonics~\cite{PRE1,JAP1}, such 
as fifth-, seventh-, etc. In doing so, we  need to keep more terms 
in powers of $\langle {\bf H}\rangle $ higher than the second in 
Eq.~(\ref{muee}). Accordingly, more terms should be included in 
Eq.~(\ref{Mor}). However, such higher-order harmonics are often 
of several orders of magnitude smaller than the third-order, and thus less attractive. 

In the numerical calculations, we have omitted the magnetizability 
of the particles due to the fact that the sizes of the 
particles are very small in ferrofluids. For these particles, 
the permanent magnetic dipole moments play the main role. 
However, in case of a magnetorheological fluid, the magnetizability 
of the particles should be taken into account since the particle 
sizes range from $2$ to $20\,$$\mu$m -- about three orders 
of magnitude larger than in ferrofluids. 
Fortunately, for treating magnetorheological fluids, the present theory holds as well.

In this paper, we have investigated a log-normal distribution [see Eq.~(\ref{lnD})]. 
Our theory could be extended to treat other particle distributions as well. 
For instance, for a $\Gamma$ distribution~\cite{gamma}, we should replace Eq.~(\ref{lnD}) with
$$
f(d)=\frac{1}{d_0}\left (\frac{d}{d_0}\right )^a\frac{e^{-d/d_0}}{\Gamma (a+1)},
$$
where $d_0$ and $a$ are the parameters of the distribution, 
and $\Gamma$ denotes the gamma function. 

In view of the existing theory which deals with the influence of 
polydispersity on the magnetization of ferrofluids, such as a perturbation 
theoretical study~\cite{r6} and a cluster expansion study~\cite{r7}, it is 
instructive to compare the present theory with these methods.

To sum up, we have used a thermodynamical method to derive the nonlinear 
magnetic susceptibility resulting from magnetostriction, which further 
yields the harmonics of magnetization in response to an applied ac magnetic field. 
For an ideal gas, this method has been shown to be in excellent agreement with a 
 statistical method.  It has been shown that the harmonics are 
sensitive to the particle distribution (namely, median diameters and standard 
deviations) and degree of  field-induced  anisotropy of the system.  In addition, we also find 
that the magnetization is higher in the polydisperse system than in the 
corresponding monodisperse one, which is in agreement with previous findings.  
Thus, it seems possible to detect the size distribution  in the polydisperse ferrofluids by    measuring the harmonics 
of the magnetization of colloidal ferrofluids under the influence of 
magnetostriction. In detail, the size distribution might be achieved by 
using Eq.~(\ref{xi-final}) and choosing  a suitable distribution for 
$f(d)$  to fit experimental data.




\section*{Acknowledgments}

We thank Prof.~H.~Pleiner of the Max Planck Institute 
for Polymer Research for a 
thorough reading of the manuscript, and Dr. M. Karttunen of the Helsinki University of Technology for carefully polishing the wording of the paper. J.\,P.\,H. acknowledges 
fruitful discussions with Prof.~K.\,W.~Yu of the Chinese
University of Hong Kong. 
This work has been supported  by
the Deutsche Forschungsgemeinschaft (German Research Foundation)  
under Grant No. HO 1108/8-3. 



\newpage


\newpage

\begin{figure}[h]
\caption{ Lognormal distribution of particles for different (a) median diameter $\delta$ and (b) standard deviation $\sigma .$ Parameters: (a) $\sigma=0.45$ and (b) $\delta=10\,$nm.  }
\end{figure}

\begin{figure}[h]
\caption{ {\it Polydisperse case.} (a) Fundamental and (b) third-order harmonics of the magnetization against the degree of anisotropy $1/g_L$ for  different median diameter $\delta .$ Parameter: $\sigma=0.45 .$ }
\end{figure}

\begin{figure}[h]
\caption{{\it Polydisperse case.}   (a) Fundamental and (b) third-order harmonics of the magnetization against the degree of anisotropy $1/g_L$ for  different standard deviation $\sigma .$  Parameter: $\delta=10\,$nm. }
\end{figure}

\begin{figure}[h]
\caption{{\it Monodisperse case.}  (a) Fundamental and (b) third-order harmonics of the magnetization against the degree of anisotropy $1/g_L$ for  different diameter $d .$ }
\end{figure}

\newpage
\centerline{\epsfig{file=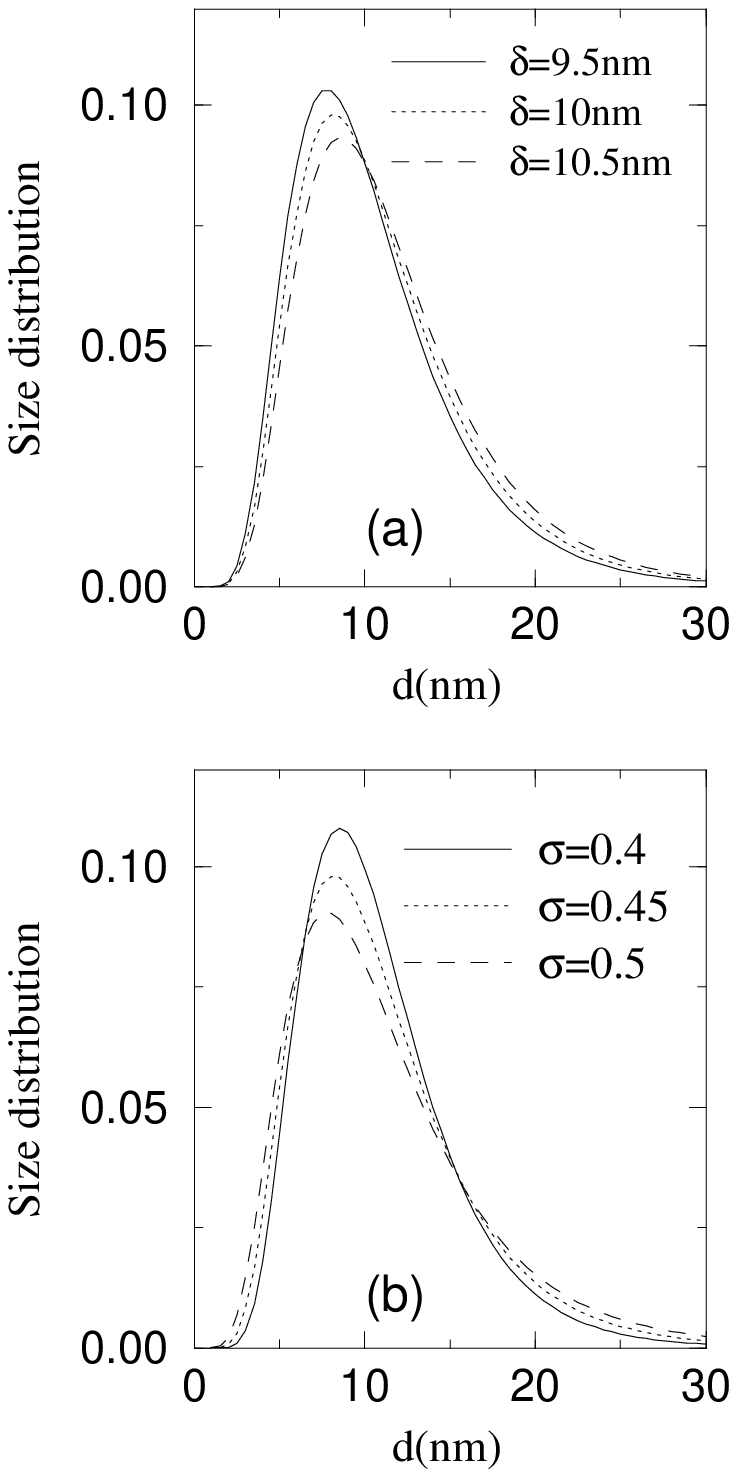,width=200pt}}
\centerline{Fig.~1/Huang and Holm}

\newpage
\centerline{\epsfig{file=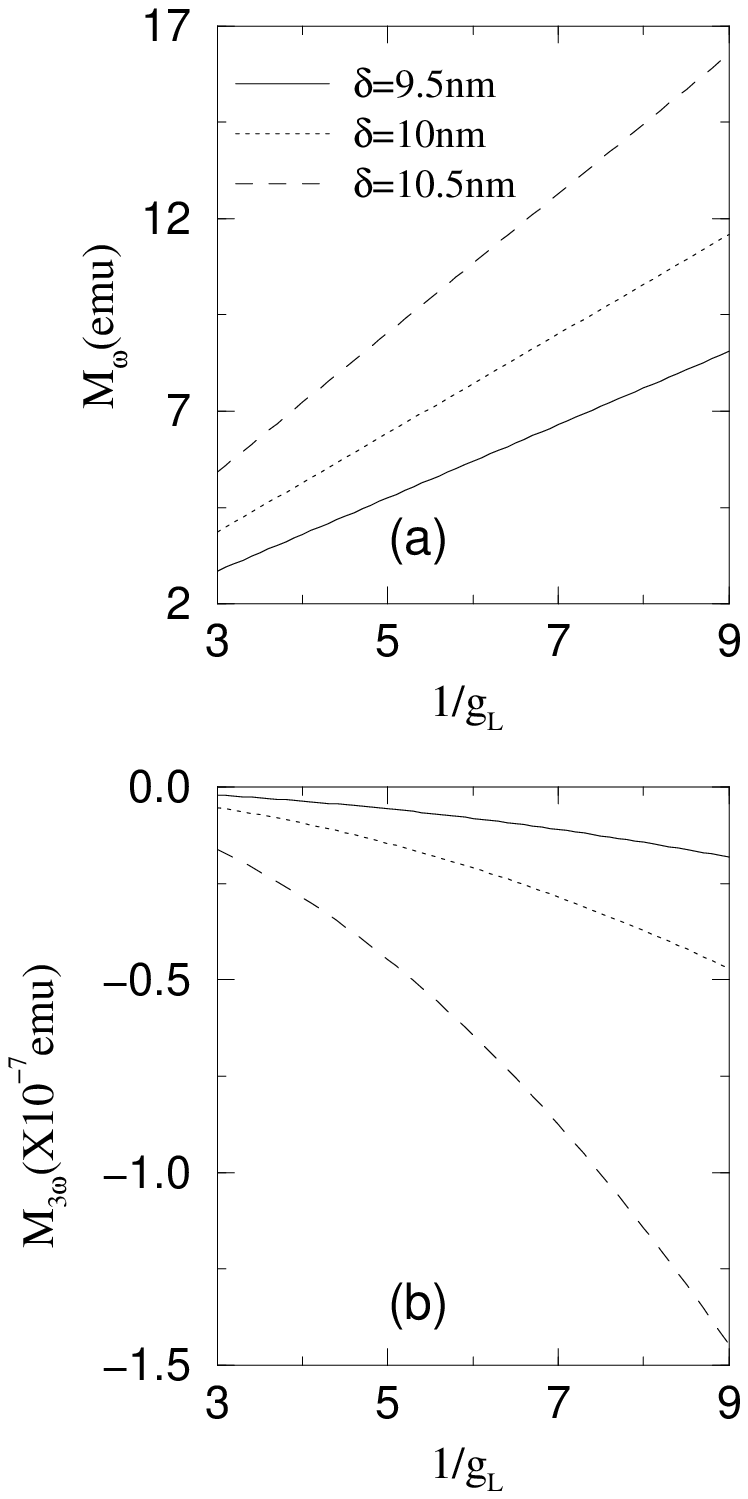,width=200pt}}
\centerline{Fig.~2/Huang and Holm}

\newpage
\centerline{\epsfig{file=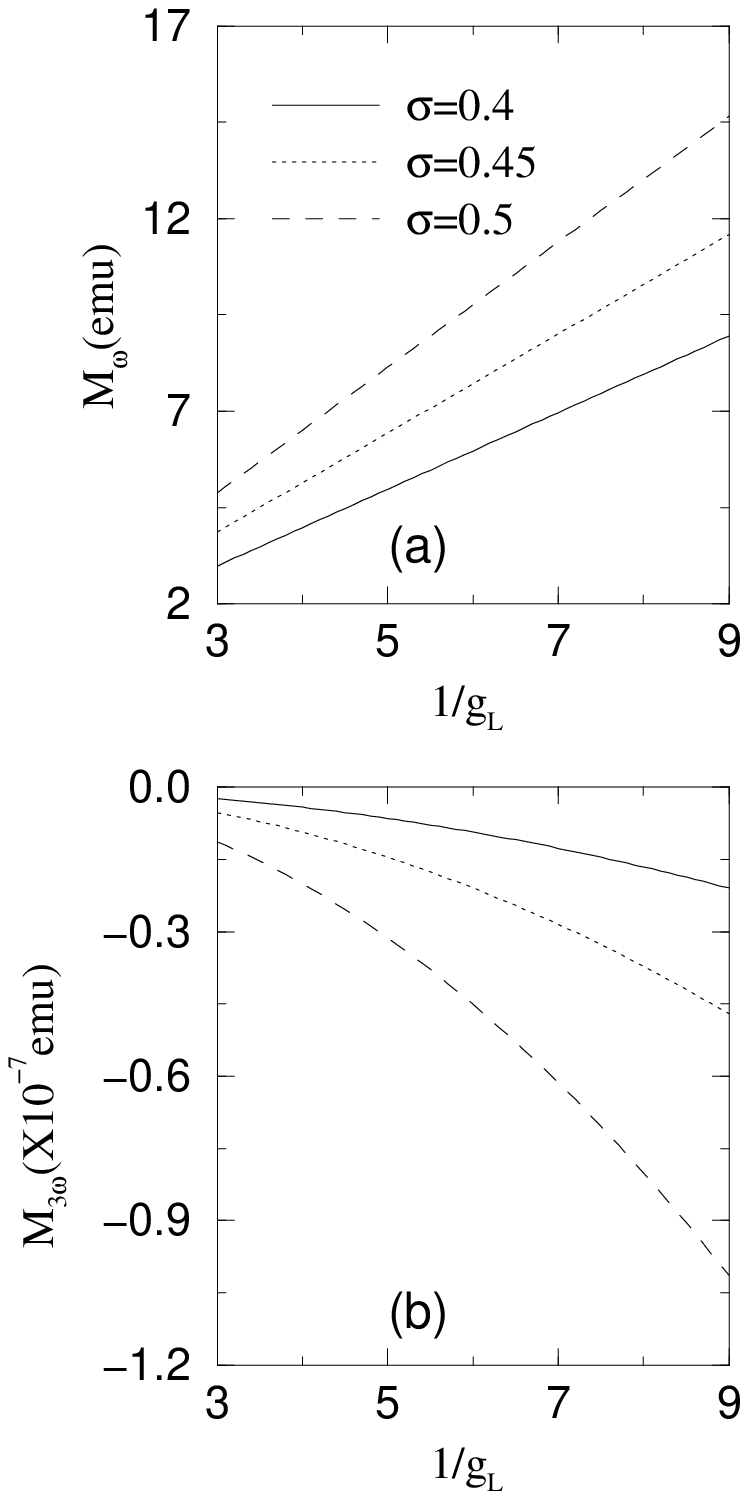,width=200pt}}
\centerline{Fig.~3/Huang and Holm}

\newpage
\centerline{\epsfig{file=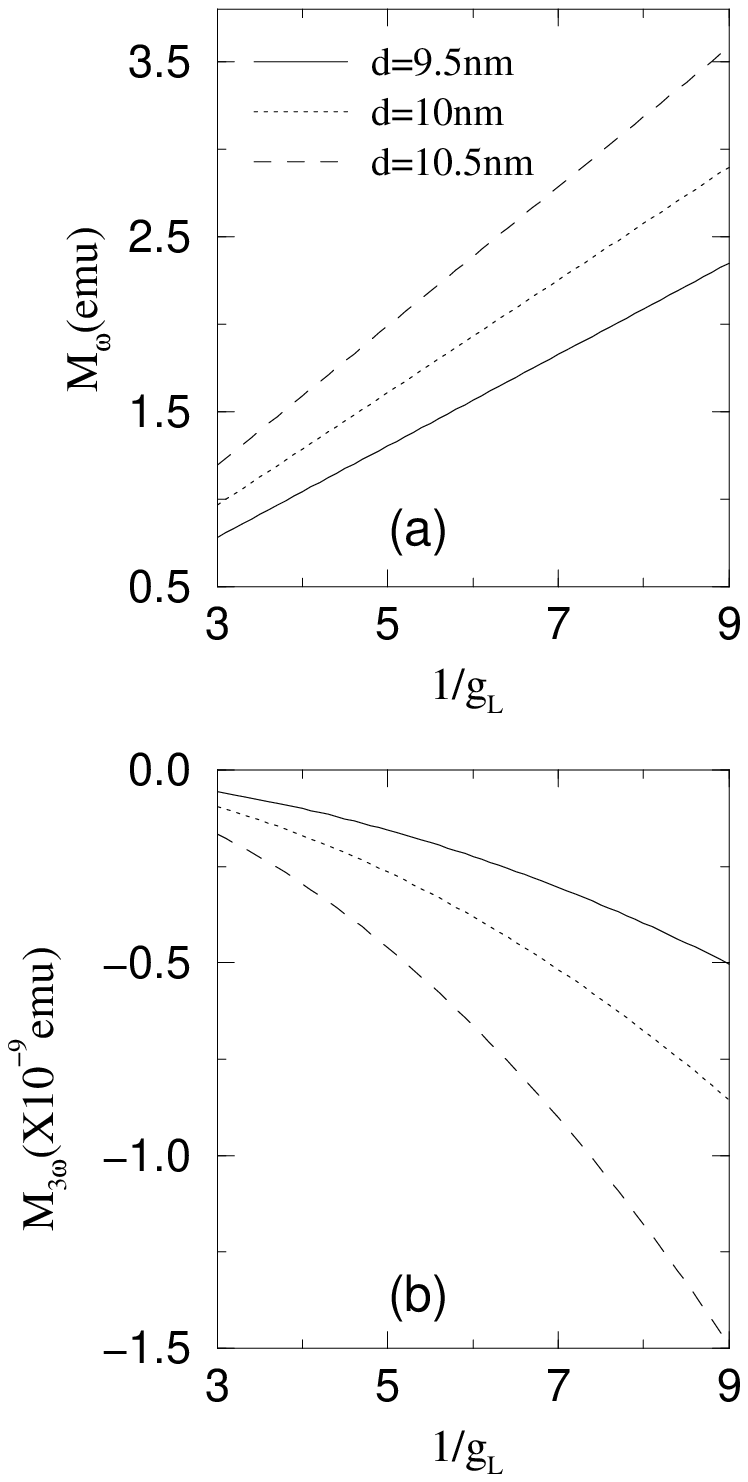,width=200pt}}
\centerline{Fig.~4/Huang and Holm}

\end{document}